\documentclass[twocolumn]{aastex631}
\usepackage{amsmath}
\usepackage{lipsum}
\usepackage{float}
\usepackage{afterpage}
\usepackage{makecell}
\usepackage{upgreek}
\usepackage{bm}
\usepackage{ulem}
\usepackage{starfont}
\usepackage[T1]{fontenc}
\usepackage{calligra}
\newcommand\as{\bgroup\markoverwith{\textcolor[rgb]{.5, 0, .6}{\rule[0.5ex]{8pt}{1.5pt}}}\ULon}

\newcommand{\be}{\begin{eqnarray}}
\newcommand{\ee}{\end{eqnarray}}

\newcommand{\RN}[1]{\textup{\uppercase\expandafter{\romannumeral#1}}}

\DeclareMathAlphabet{\mathcalligra}{T1}{calligra}{m}{n}
\DeclareFontShape{T1}{calligra}{m}{n}{<->s*[2.2]callig15}{}

\graphicspath{{./}{figures/}}
\shorttitle{}
\shortauthors{Dittmann, Dempsey, \& Li}
\begin{document}
\title{The Multiple Paths to Merger of Unequal-Mass Black Hole Binaries\\ in the Disks of Active Galactic Nuclei}
\shorttitle{Unequal-Mass Binaries in AGN Disks}

\correspondingauthor{Alexander J. Dittmann}
\email{dittmann@ias.edu}

\author[0000-0001-6157-6722]{Alexander J. Dittmann}
\affil{Institute for Advanced Study, 1 Einstein Drive, Princeton, NJ 08540, USA}
\affiliation{NASA Hubble Fellowship Program, Einstein Fellow}

\author[0000-0001-8291-2625]{Adam M. Dempsey}
\affil{X-Computational Physics Division, Los Alamos National Laboratory, Los Alamos, NM 87545, USA}

\author[0000-0003-3556-6568]{Hui Li}
\affil{Theoretical Division, Los Alamos National Laboratory, Los Alamos, NM 87545, USA}

\begin{abstract} 
The accretion disks that power active galactic nuclei (AGN) are thought to house populations of stars and compact objects; after forming binaries these compact objects may merge, begetting gravitational waves such as those detected by LIGO and VIRGO. We present a comprehensive study of the early evolution of binaries within AGN disks as their orbits are influenced by the surrounding gas, focusing on eccentric and unequal-mass binaries. Nearly-equal-mass binaries behave similarly to their equal-mass counterparts: prograde binaries inspiral, albeit somewhat slowly, and have their eccentricities damped; retrograde binaries inspiral $\sim2-3$ times faster than their prograde counterparts, and those with near-equal masses are driven quickly towards near-unity eccentricities. However, the primaries in retrograde binaries with mass ratios of $m_2/m_1\lesssim0.4$ experience significantly weaker headwinds and retain substantial accretion disks that help damp binary eccentricities, slowing binary inspirals. Additionally, we find that while accretion drives prograde binaries towards equal masses thanks to the exchange of material between the primary and secondary accretion disks, retrograde binaries are driven slowly towards more extreme mass ratios. Prograde binaries, and generally those with low mass ratios, likely accrete for multiple $e$-folding timescales before merger. On the other hand, high-mass-ratio retrograde binaries may merge before accreting substantially, potentially approaching merger with detectable eccentricity. Future ground-based gravitational wave observatories, with their broader frequency coverage, should be particularly useful for studying these populations.
\end{abstract}


\section{Introduction}
The accretion disks around supermassive black holes (SMBHs) that power active galactic nuclei \citep[AGN; e.g.,][]{1969Natur.223..690L,2008ARA&A..46..475H} may also host numerous stars and compact objects, including those captured from nuclear star clusters \citep[e.g.,][]{1993ApJ...409..592A,2020MNRAS.499.2608F} or formed in situ \citep[e.g.][]{1980SvAL....6..357K,2024OJAp....7E..71H}. The deep potential wells of these SMBHs retain stellar-mass black hole binary (BHB) merger products despite gravitational wave-induced kicks, facilitating multiple generations of mergers; the ample supply of gas can also feed embedded black holes at near or above the Eddington limit \citep[e.g.,][]{2012MNRAS.425..460M,2021NatAs...5..749G}. AGN disks can thus readily beget high-mass BHB mergers such as the one that caused GW190521, a gravitational wave (GW) event that resulted from the merger of black holes with masses $85_{-14}^{+21}\,M_\odot$ and $66_{-18}^{+17}\,M_\odot$ at 90\% confidence, (\citealt{2020PhRvL.125j1102A}; but see \citealt{2021ApJ...907L...9N}). Although the presence of gas around black hole binaries introduces the \textit{possibility} of electromagnetic counterparts to their mergers \citep[e.g.][]{2017MNRAS.464..946S,2019ApJ...884L..50M,2023ApJ...950...13T}, attempts to associate flaring AGNs with GW events have been largely inconclusive, albeit with a number of potential associations \citep[e.g.][]{2021ApJ...914L..34P,2023ApJ...942...99G,2025MNRAS.536.3112V}.\footnote{It is also worth noting that at fairly high confidence, based on the spatial correlations (or lack thereof) between GW event error volumes and the locations of known AGN, fewer than roughly 50\% of the BHB mergers detected thus far can originate from nearby AGN with $z\leq0.3$ and bolometric luminosities greater than $10^{45.5}$ erg s$^{-1}$ \citep{2023MNRAS.526.6031V}. However, there are also hints that a substantial fraction of GW events may be associated with lower-luminosity AGN \citep[e.g.,][]{2025arXiv250502924Z}}

Although models, from simple semi-analytical calculations to radiation magnetohydrodynamic accretion disk simulations, predict that AGN disks should host hundreds or more stars and compact objects \citep[e.g.,][]{2022ApJ...928..191G,2024OJAp....7E..71H}, the connections between those populations and the black hole mergers detected by LIGO, Virgo, and Kagra (LVK) remain highly uncertain. Although progress has been made towards modeling these populations \citep[e.g.,][]{2024arXiv241016515M,2025ApJ...979..245D}, the path to merger from binary formation is highly uncertain: not only do binaries typically form with semi-major axes of some modest fraction of their Hill radii, often $\sim7$ orders of magnitude larger than their gravitational radii \citep[see, for example, Section 2 of][]{2022ApJ...940..155D}, but hydrodynamical interactions may even cause binaries to expand \citep[e.g.,][]{2017MNRAS.466.1170M,2024ApJ...970..156D}, potentially causing the binary to become dynamically unstable \citep[e.g.,][]{2001MNRAS.321..398M} or even unbound.

Over the last few years it has become possible to simulate both the formation and subsequent orbital evolution of binaries with substantially increased fidelity. Recent studies have shed light on the role of various processes in binary formation, both dynamical \citep{2022ApJ...934..154L,2023MNRAS.521..866R,2024ApJ...962..143Q} and hydrodynamical \citep{2023ApJ...944L..42L,2023MNRAS.524.2770R,2024MNRAS.531.4656W,2025arXiv250214959W}. The aforementioned studies have illustrated how binaries can form in AGN disks with a wide range of eccentricities and in both prograde and retrograde configurations (with respect to the center-of-mass motion of the binary). 

Concurrently, high-resolution local \citep{2022MNRAS.517.1602L,2023MNRAS.522.1881L,2024MNRAS.529..348L} and global \citep{2021ApJ...911..124L,2022ApJ...928L..19L} two-dimensional simulations have surveyed a wide variety of disk condition and binary parameters at the neglect of three dimensional effects. Three-dimensional simulations, which are able to capture vertical flows of gas that strongly affect binary evolution, have been more limited in scope, but thus far explored disk thermodynamics and binary separation \citep{2022ApJ...940..155D}, the realignment of out-of-plane binaries \citep{2024ApJ...964...61D}, and both prograde and retrograde eccentric binaries \citep{2024ApJ...970..107C}.\footnote{A few three dimensional wind-tunnel studies have also been conducted \citep{2023ApJ...944...44K,2024ApJ...977...16R}, although their neglect of Coriolis terms limits their applicability to binaries within accretion disks.} Intriguingly, the drag forces experienced by retrograde binaries tend to drive them towards eccentricities of unity, potentially facilitating relatively rapid mergers. However, three-dimensional simulations have thus far focused on equal-mass binaries; because the circumbinary dynamics observed in near-equal-mass binaries must break down at extreme mass ratios (as the system reduces to a single point mass embedded within the disk), we aim to rectify this limitation of prior studies. 

We present herein an investigation into the orbital evolution of eccentric, unequal-mass binaries embedded within AGN disks. %\footnote{And of course, exomoons.}
Section \ref{sec:methods} describes our numerical methodology and Section \ref{sec:results} presents the results of our simulations. We discuss some of the implications and limitations of our findings in Section \ref{sec:discussion}, and conclude in Section \ref{sec:conclusion}. Appendix \ref{sec:appendix} tabulates our simulation results.

\section{Methods}\label{sec:methods}
To study the orbital evolution of black hole binaries embedded within AGN disks, we have employed a modified version of the \texttt{Athena++} code \citep{2020ApJS..249....4S}, coupled with the \texttt{REBOUND} $N$-body integrator \citep{2012A&A...537A.128R} as described in Section 3.1 of \citet{2024ApJ...964...61D}. We have assumed that the gas is isothermal and inviscid, and have used the shearing-box approximation \citep{1965MNRAS.130..125G,1996ApJ...463..656S}, expanding the equations of hydrodynamics about the initial center of mass position of the embedded binary. This method enables our simulations to capture global features such as the excitation of spiral arms in the AGN disk while resolving the three-dimensional flow of gas around the binary with high fidelity. 

\subsection{The embedded binary}
We restrict ourselves in this work to binaries either aligned (prograde, having an orbital angular momentum vector parallel to that of the binary about the SMBH) or anti-aligned (retrograde, having an orbital angular momentum vector parallel to that of the binary about the SMBH) with the orbital plane of the binary through the AGN disk, which we also take to be the AGN disk midplane. We only consider binaries with prograde center-of-mass motion through the AGN disk. Restricting our survey of binary inclinations has enabled us to more broadly survey other binary parameters, such as their mass ratios and eccentricities. In the following, we summarize some of the definitions and relationships pertinent to this study, but see \citet{2024ApJ...964...61D} for illustrations and a more general discussion.

Each binary is characterized by its specific energy and angular momentum, 
\begin{equation}
\mathcal{E}=\frac{1}{2}\mathbf{v}\cdot\mathbf{v}-\frac{Gm}{r},~~~~\mathbf{h}=\mathbf{r}\times\mathbf{v},
\end{equation}
where $m=m_1+m_2$ is the total mass of the embedded binary, $\mathbf{r}=\mathbf{r}_2-\mathbf{r}_1$ is the relative position vector and $\mathbf{v}=\mathbf{v}_2-\mathbf{v}_1$ is the relative velocity vector. The binary eccentricity and semi-major axis can be defined in terms of $\mathcal{E}$ and $h\equiv \sqrt{\mathbf{h}\cdot\mathbf{h}}$ according to
\begin{equation}
a=-\frac{Gm}{2\mathcal{E}},~~~e^2=1-\frac{h^2}{Gma}.
\end{equation}
The orientation of the binary within its orbital plane is defined by its eccentricity vector, which points towards periapse, 
\begin{equation}
\mathbf{e}=\frac{\mathbf{v}\times\mathbf{h}}{Gm}-\frac{\mathbf{r}}{r},
\end{equation}
where $r\equiv\sqrt{\mathbf{r}\cdot\mathbf{r}}$.
The orientation of the orbit is described by its longitude of pericenter $\varpi$, which relates the components of the eccentricity vector to its magnitude
\begin{equation}
e_x = e\cos{\varpi},~~e_y=e\sin{\varpi}.
\end{equation}

The specific energy and angular momentum of the binary are related to the total energy ($E$) and angular momentum ($J$) of the binary through the reduced mass ($\mu$)
\begin{equation}\label{eq:EJ}
E=\mu\mathcal{E},~~~J=\mu h
\end{equation}
where the reduced mass is defined as 
\begin{equation}
\mu=\frac{m_1m_2}{m}=\frac{qm}{(1+q)^2}
\end{equation}
and the mass ratio $q\equiv m_2/m_1$. 

One should note that, since 
\begin{equation}
\frac{\dot{q}}{q} = \frac{\dot{m}_2}{m_2} - \frac{\dot{m}_1}{m_1},
\end{equation}
the condition that the binary mass ratio increases due to accretion can be expressed as 
\begin{equation}
\frac{\dot{m}_2}{\dot{m}_1} > \frac{m_2}{m_1}.
\end{equation}
Thus, unless $\dot{m}_2/\dot{m}_1<q$, the binary will be driven towards a mass ratio of unity.

The rates of change of the binary semi major axis and eccentricity can be inferred by measuring the torque and power exerted on the binary by its interactions, through gravity and accretion, with its environment. These expressions follow from Equation \ref{eq:EJ} and are given by
\begin{align}
\frac{\dot{a}}{a} &= \frac{\dot{m}_1}{m_1} + \frac{\dot{m}_2}{m_2} - \frac{\dot{E}}{E},\\
\frac{e\dot{e}}{1-e^2}&= \frac{\dot{m}}{m} + \frac{3\dot{\mu}}{2\mu} - \frac{\dot{E}}{2E} - \frac{\dot{J}}{J}.
\end{align}
A detailed description of how we measure these quantities from our simulations is provided by Appendix A of \citet{2024ApJ...964...61D}. 

We consider binaries, as described above, in hierarchical triple systems with SMBHs of mass $M_\bullet$, where $m/M_\bullet \ll1$. The center of mass of the binary is initialized in a circular orbit about the SMBH with angular frequency $\Omega_0$ and radius $R_0 \gg a$. It is worth noting that the orbital elements of the binary are not fixed in time, but vary due to the tidal potential of the SMBH. We evolve the orbital evolution of this triple system self-consistently using direct 3-body simulations \citep{2012A&A...537A.128R,2015MNRAS.446.1424R}, which causes the orbital elements of the binary to fluctuate over time \citep[see, for example, Figure 1 of][]{2024ApJ...970..107C}.

\subsection{Hydrodynamics}
Our model for the fluid surrounding the BHB is described by the equations
\begin{equation}
\partial_t\rho + \nabla\cdot(\rho\mathbf{v})=S_\rho,
\end{equation}
\begin{multline}
%\begin{split}
\partial_t(\rho\mathbf{v})\!= - \!\nabla\!\cdot\!(\rho\mathbf{v}\mathbf{v}+P) -\rho\nabla\Phi+\mathbf{S}_p\\+\rho\Omega_0^2(3\hat{\mathbf{x}}-\hat{\mathbf{z}}) - 2\rho\Omega_0\hat{\mathbf{z}}\times\mathbf{v},
%\end{split}
\end{multline}
%\begin{equation}
%P=c_s^2\rho,
%\end{equation}
where $\Phi$ is the gravitational potential of the BHB, $S_\rho$ is a mass sink term, $\mathbf{S}_p$ is a momentum sink term, $c_s$ is a globally-constant sound speed, and we have assumed a Keplerian shear flow. The angular velocity of the simulation domain is set to the Keplerian value, $\Omega_0=\sqrt{GM_\bullet/R_0^3}.$
The coordinate basis vectors in this frame rotate as the binary orbits about the central SMBH, and are $\hat{\mathbf{x}}$, which points from the SMBH to the center of the shearing box, $\hat{\mathbf{z}}$, which is normal to the AGN disk midplane, and $\hat{\mathbf{y}}=\hat{\mathbf{z}}\times\hat{\mathbf{x}}$. We chose the sound speed so that the scale height of the disk, defined as $H_0=c_s\Omega_0^{-1}$, is given by $H_0=R_0/100$, a typical value for AGN disk models \citep[e.g.][]{2002apa..book.....F}.

We modeled accretion onto each point mass using the mass and momentum sink terms $S_\rho$ and $\mathbf{S}_p$, which were localized within a region or radius $r_s$ around each point mass. The momentum sink term was tuned to preclude artificial torques on the gas, which have the potential to malignantly affect the larger-scale accretion flow \citep{2020ApJ...892L..29D,2021ApJ...921...71D}, especially in unequal-mass binaries \citep[see][Appendix A]{2024ApJ...967...12D}. We modeled the gravitational potential of the binary, $\Phi$ as the sum of two Newtonian potentials, each softened to avoid singularities while recovering the exact Newtonian potential outside of the softening radius \citep{2001NewA....6...79S}, which was chosen to be equal to the sink radius and set to $r_s=0.04a.$ Our simulations were initialized using an unperturbed disk in hydrostatic equilibrium, and the binary was smoothly introduced between $t=0$ and $t=0.5\Omega_0^{-1}$.

The domains of our simulations extended across $[-24H_0,24H_0]\times[-24H_0,24H_0]\times[-4H_0,4H_0]$ in the $\hat{\mathbf{x}}$, $\hat{\mathbf{y}}$, $\hat{\mathbf{z}}$ directions respectively. We simulated binaries with outer mass ratio $m/M_\bullet=1.536\times10^{-6}$, such that their Hill radii \citep{hill1878researches}
\begin{equation}
R_H=R_0\left(\frac{m}{3M_\bullet}\right)^{1/3}
\end{equation}
were $R_H=0.8H_0$, and with semi-major axes of $a=0.25R_H$ so as to be dynamically stable \citep{2001MNRAS.321..398M}.
Our base grid resolved the $H_0$ using 4 cells in each direction, and we employed seven levels of static mesh refinement to resolve the binary semi-major axis by 102 cells. We carried out a limited set of runs at $q=0.3,\,e=0.5$ using twice the spatial resolution in each dimension and found time-averaged results in agreement with those from our standard-resolution runs.\footnote{In Appendix \ref{sec:appendix}, we also make a posteriori estimates of the resolution-dependent errors in our simulations, which suggest that the values measured in our simulations may be accurate \textit{in terms of truncation error} to a few percent.} Our simulations employed 2nd-order spatial reconstruction \citep{1974JCoPh..14..361V}, the HLLE approximate Riemann solver \citep{10.2307/2030019,1988SJNA...25..294E}, and a second-order strong stability preserving Runge-Kutta method for evolution in time \citep[RK2][]{heun1900neue,1998MaCom..67...73G}.
We used outflow boundary conditions in the $x$ direction and periodic boundary conditions in the $y$ direction. We applied outflow boundary conditions to the velocities in the $z$ direction, but extrapolated the density so as to maintain vertical hydrostatic equilibrium.

\begin{figure}
\includegraphics[width=\columnwidth]{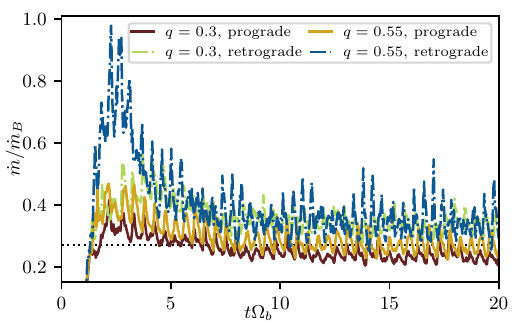}
\caption{The rates of accretion, normalized by the Bondi accretion rate, onto both prograde and retrograde binaries with $e=0.5$ and mass ratios $q=0.3$ and $q=0.55$. All the binaries accrete at about the rate expected based on their Hill radii ($\dot{m}/\dot{m}_B\sim(R_H/R_B)^2=0.2713$; dotted line), and retrograde binaries accrete at slightly higher rates than prograde ones, as discussed in \citet{2024ApJ...964...61D}. In all cases, the accretion rates reach a quasi-steady state by $t\sim8\Omega_b^{-1}$.}
\label{fig:accTimeSeries}
\end{figure}
\begin{figure*}
\includegraphics[width=\linewidth]{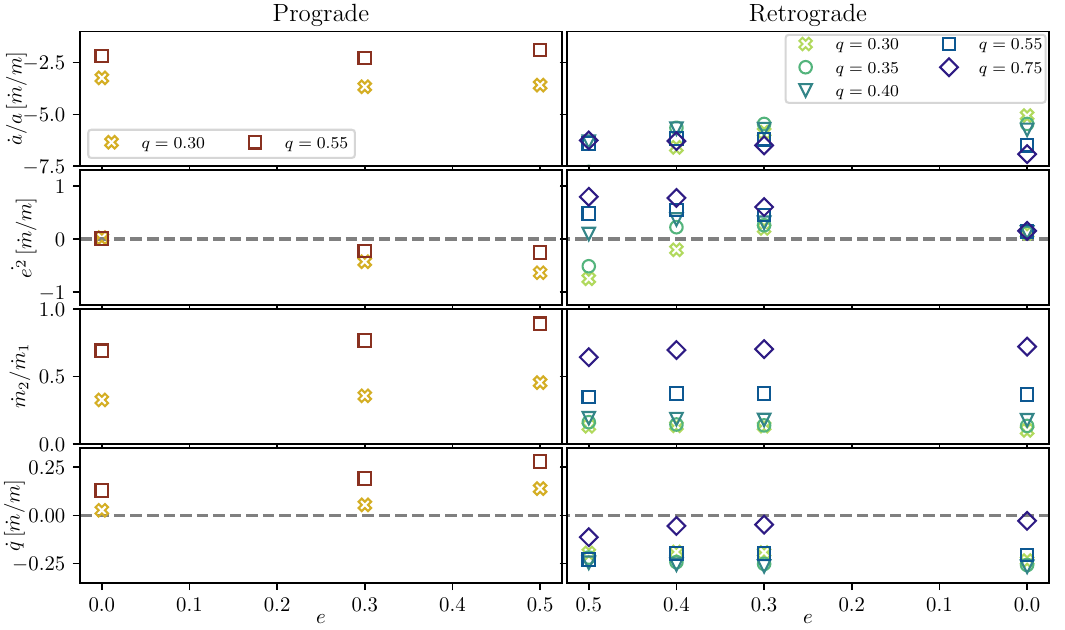}
\caption{A summary of binary evolution for the eccentricities and mass ratios studied in this paper, grouping prograde binaries on the left-hand side and retrograde binaries on the right. Each time derivative has been expressed in terms of the time-averaged binary accretion rate and binary mass, $\dot{m}/m$. The top row displays the average rate of change of the binary semi-major axis due to accretion and gravitational interactions with their surroundings, normalized by the time-averaged binary semi-major axis. The second row displays the time-averaged rate of change of the binary eccentricity squared, $\dot{e^2}=2e\dot{e}$, which we plot here as it is well-defined even for circular binaries. 
The third and fourth rows display the ratio of the time-averaged accretion rate onto the secondary divided by that onto the primary, and the bottom panel plots the rate of change of the binary semi-major axis in each simulation.
The dashed gray lines indicate the boundaries between eccentricity pumping and damping in the second row, and the division between equalization and disequalization of the binary masses in the fourth panel.
}
\label{fig:scatter}
\end{figure*}
Each simulation was run until at least $t=20\Omega_0^{-1}$, a bit more than three orbital periods of the binary about the SMBH and more than forty orbital periods of the binary itself. The flow of gas onto the binary typically reached a quasi-steady state by $t\sim8\Omega_0^{-1}$, as illustrated by the accretion rate times series in Figure \ref{fig:accTimeSeries}. In a few cases, specifically retrograde binaries with $e=0.5$, we continued each simulation until $t=100\Omega_0^{-1}$ to improve our orbital-phase-resolved analyses. We used in-situ per-timestep records of the various forces (from both accretion and gravity) on each member of the binary to infer the time-evolution of its orbital elements as described in Appendix A of \citet{2024ApJ...964...61D}.

\section{Results}\label{sec:results}

We surveyed binary-disk interactions over a broad range of binary parameters. 
We initially simulated both prograde and retrograde binaries with mass ratios $q\in\{0.3, 0.55\}$ and eccentricities $e\in\{0, 0.3, 0.5\}$. After finding that the orbital evolution of retrograde binaries depended more sensitively on both eccentricity and mass ratio, we additionally studied retrograde binaries at mass ratios $q\in\{0.35, 0.4, 0.75\}$ and eccentricities $e=0.4$, simulating binaries with each combination of $q$ and $e$.

\begin{figure}
\includegraphics[width=\linewidth]{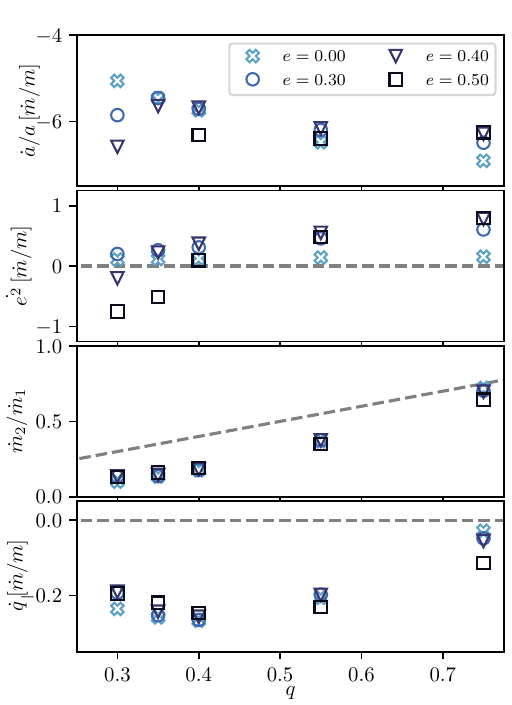}
\caption{A summary of binary evolution for the retrograde binaries studied in this paper as a function of binary mass ratio, as in Figure \ref{fig:scatter}. In the second panel, the dashed line demarcates the boundary between eccentricity damping and excitation. In the third and fourth panels, the dashed gray lines indicate the $\dot{q} = 0$ boundary.}
\label{fig:qscatter}
\end{figure}

Our binary orbital evolution results are summarized in Figure \ref{fig:scatter} as a function of binary eccentricity, and for retrograde binaries as a function of mass ratio in Figure \ref{fig:qscatter}. 
In all cases, we found that the primary accretes at a higher rate than the secondary; that is to say, that the more massive object in the binary accretes preferentially. In the prograde cases, $\dot{m}_2/\dot{m}_1>q$ so that $\dot{q}>0$ and binaries will be driven towards equal masses over time; however, in the retrograde cases the accretion rate onto the secondary is low enough that mass ratios actually become more extreme over time. These trends have only minor dependence on the binary eccentricity, and are almost entirely a consequence of the binary mass ratio. We show in Section \ref{sec:accretion} how these follow from the predominantly vertical nature of accretion in these systems and the transfer of matter (or lack thereof) from one member of the binary to the other. 

Previous three-dimensional simulations \citep{2024ApJ...964...61D,2024ApJ...970..107C} found that retrograde binaries inspiral more quickly than prograde binaries due to binary-disk interactions, and that while the eccentricities of prograde binaries are damped, the eccentricities of retrograde binaries should be driven towards unity. We confirm most of these results in Figures \ref{fig:scatter} and \ref{fig:qscatter}, with one exception: retrograde binaries with mass ratios below $q\lesssim0.4$ have their eccentricities damped rather than excited, driving binaries towards an equilibrium near $e\sim0.4$. As we illustrate in Section \ref{sec:eccentricity}, this behavior occurs because the primaries of binaries with $q\lesssim0.4$ are able to retain their own accretion disks (minidisks); at eccentricities above $e\gtrsim0.3$ the secondaries in these low-mass ratio systems begin to strongly interact with the minidisk of the primary near pericenter, damping eccentricity faster than it can be excited near apocenter.

\subsection{Accretion}\label{sec:accretion}
The total accretion rate onto a given binary is very weakly dependent on the binary mass ratio and eccentricity, but strongly dependent the binary orientation, as illustrated in Figure \ref{fig:accTimeSeries} and in Table \ref{tab:orbit}. The relative accretion rates onto the primary and secondary depend on the binary orientation and mass ratio, but not the binary eccentricity, as shown in Figures \ref{fig:scatter} and \ref{fig:qscatter}. Notably the accretion rate onto the binaries quite closely matches the prediction of $\dot{m}=(R_H/R_B)^2\dot{m}_B$, where $R_B=Gm/c_s^2$ is the Bondi radius and $\dot{m}_B=\pi c_s R_B^2$ \citep{1952MNRAS.112..195B}.\footnote{In slightly more detail, the Bondi accretion rate must be modified once the Bondi radius of a binary becomes larger than its Hill radius since beyond that point gas would no longer be bound to the the binary. See, for example, \citet{2020MNRAS.498.2054R,2021ApJ...916...48D,2023MNRAS.525.2806C}.}

\begin{figure}
\includegraphics[width=\columnwidth]{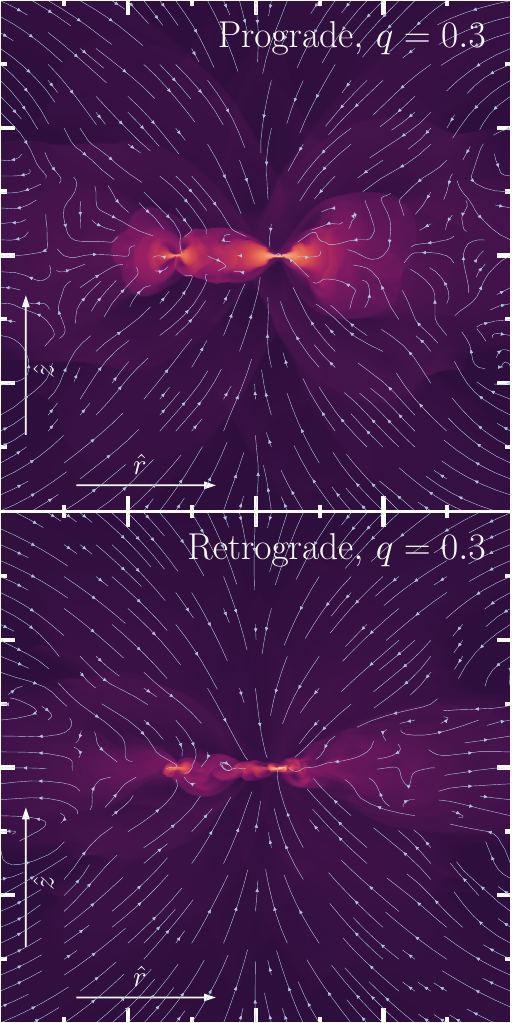}
\caption{Slices of the gas density along the binary separation vector for a pair of $q=0.3,\,e=0$ binaries, centered on the center of mass of each binary. At larger scales gas falls ballistically towards the center of mass of the binary; at smaller scales, most of this material is captured by the primary thanks to its extensive gravitational influence, although the secondary manages to captures some as well. This follows from large-scale gas dynamics, and occurs regardless of whether a given black hole possesses a robust minidisk, as illustrated by the similarity between the prograde and retrograde cases. Similarly, the partition of accretion onto the primary is largely independent of binary eccentricity. Major ticks are placed every $1.25\,a$ and minor ticks are placed every $0.675\,a$.}
\label{fig:slices}
\end{figure}

That the primary tends to accrete preferentially follows from the predominantly vertical nature of accretion in these systems. Although studies of \textit{thin} circumbinary disks typically find that the secondary object accretes preferentially \citep[e.g.,][]{2014ApJ...783..134F,2020ApJ...889..114M,2024ApJ...967...12D}, that result is a quasi-2-dimensional peculiarity caused by the vertical confinement of these accretion flows. Binaries embedded in AGN disks on the other hand, similarly to planets \citep[e.g.,][]{2014ApJ...782...65S,2016ApJ...832..105F}, can accrete primarily in the vertical direction. This has been illustrated on quasi-global scales in Figure 4 of \citet{2022ApJ...940..155D} and Figure 6 of \citet{2024ApJ...964...61D}. The picture changes somewhat as gas approaches the binary: as illustrated in Figure \ref{fig:slices}, as gas approaches the binary it is redirected from the center of mass of the binary towards one of the objects. Since the primary has much more extensive gravitational influence, it captures a majority of the vertically infalling material. As illustrated by Figure 2 of \citet{2024ApJ...970..107C}, infalling material is equally divided between the members of equal-mass binaries. In the unequal-mass case, however, the primary manages to capture more material in both prograde and retrograde binaries. 

\begin{figure}
\includegraphics[width=\columnwidth]{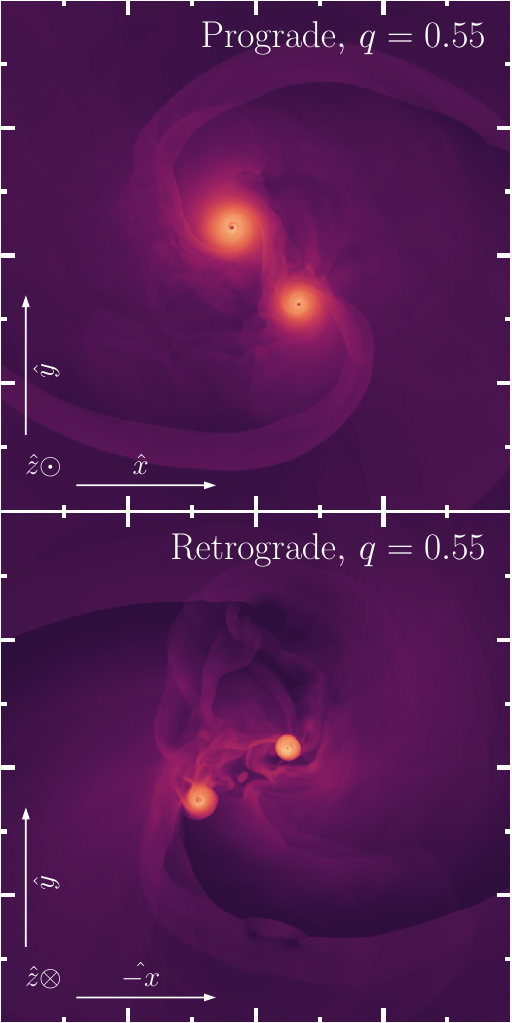}
\caption{The surface density of gas in the orbital plane of two $q=0.55,\,e=0$ binaries, centered on the center of mass of each binary. Since the sizes of the minidisks around the black holes in prograde binaries are comparable in size to the Roche lobes of those black holes, matter readily flows between the two members of the binary, which tends to help the secondary accrete material initially captured by the primary. Such transfer is reduced in retrograde binaries. Major ticks are placed every $1.25\,a$ and minor ticks are placed every $0.675\,a$.}
\label{fig:slicesXY}
\end{figure}

The preferential gravitational capture of material by the primary follows naturally from the greater extent of its basin of attraction and the largely vertical nature or accretion. While this explains the general preference for accretion onto the primary, it fails to explain why the secondary tends to accrete even less in retrograde binaries than prograde ones, and thus why $\dot{q}>0$ for prograde binaries but $\dot{q}<0$ for retrograde ones. The crucial difference that causes this discrepancy is that prograde binaries are able to sustain robust minidisks, while the accretion disks around retrograde binaries are stripped away by ram pressure. As illustrated in Figure \ref{fig:slicesXY}, streams of gas flow between the members of prograde binaries, exchanging material between them. Such streams are absent in retrograde systems, leaving fewer mechanisms to rebalance accretion between primary and secondary. 

\begin{figure}
\includegraphics[width=\columnwidth]{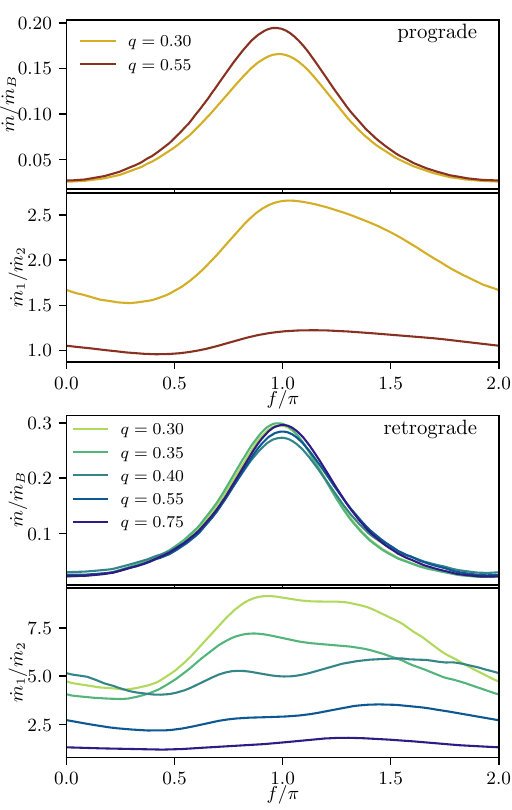}
\caption{The accretion rate (first and third rows, here weighted by the amount of time spent at each phase) and ratio of primary to secondary accretion rates (second and fourth rows) onto $e=0.5$ binaries as functions of orbital phase. 
In all cases the binaries primarily accrete near apocenter ($f=\pi$), where they spend the most of their orbit. Very little mass is exchanged between the primary and secondary in high-mass-ratio retrograde binaries, and thus $\dot{m}_1/\dot{m}_2$ is roughly constant over time. On the other hand, prograde and lower-mass-ratio retrograde binaries, which support substantial accretion flows around their primaries, transfer considerable matter from primary to secondary near pericenter, decreasing $\dot{m}_1/\dot{m}_2$ at those orbital phases. }
\label{fig:mdotfolded}
\end{figure}

The role of streams in exchanging matter between the Roche lobes of primary and secondary members has been studied previously in the context of isolated circumbinary accretion: in those systems the secondary tends to accrete at a higher rate thanks to the planarity of the system, but streams of material transfer material from the secondary to the primary. For example, \citet{2015MNRAS.447.2907Y} determined that in relatively hot, puffy but still thin accretion disks, the much stronger influence of pressure gradients within the Roche lobe of the secondary enhanced the transfer of material through the L1 Lagrange point, explaining earlier simulations that found preferential accretion onto the primary \citep[e.g.,][]{2005ApJ...623..922O,2010ApJ...708..485H} whereas simulations of accretion onto thinner disks tended to find preferential accretion onto the secondary \citep[e.g.,][]{1997MNRAS.285...33B}. Similarly, the spatial profile of the eddy viscosity employed in 2D vertically-integrated simulations (e.g. $\alpha$ vs constant-$\nu$) can also enhance the amount of matter flowing from secondary to primary by factors of $\gtrsim3$ \citep{2024ApJ...967...12D}. Thus, it is not particularly surprising that the absence of flows between the minidisks of these embedded binaries should cause the primaries to accrete at substantially higher rates in prograde binaries, or for $\dot{q}$ to be negative in those systems.

To illustrate this process more quantitatively, Figure \ref{fig:mdotfolded} plots the orbital-phase-averaged accretion rate onto a collection of prograde and retrograde $e=0.5$ binaries, as well as the ratio of the phase-averaged accretion rates onto the secondary and primary. In each case the total accretion rate is highest near apocenter ($f=\pi$), where binaries spend the largest temporal fraction of their orbit, and the total accretion rate only weakly depends on the binary mass ratio. However, the partition of accretion between the primary and secondary varies significantly over phase and between mass ratios. In prograde binaries, the ratio of $\dot{m}_1/\dot{m}_2$ is greatest at apocenter, where the the binary components have minimal mutual interaction, and lowest near pericenter where the secondary is able to capture material from the primary. Higher-mass ratio retrograde binaries show much less variation over orbital phase, since each minidisk only occupies a small fraction of its corresponding Roche lobe. \textit{At apocenter} there is a monotonic trend in $\dot{m}_1/\dot{m}_2$, which increases in retrograde binaries as the primary becomes more massive. At higher mass ratios, $q\gtrsim0.4$, this trend holds across orbital phases and $\dot{m}_1/\dot{m}_2$ is roughly constant in orbital phase; in lower-mass-ratio binaries however, trends similar to those observed in prograde binaries occur, as matter initially captured by the primary is transferred. This change occurs because at sufficiently low mass ratios, retrograde primaries experience low enough ram pressure stripping to sustain minidisks, which will be discussed in more detail in the following section and is shown in Figure \ref{fig:slicePeri}. This transfer of material between primary and secondary is responsible for the nonmonotonicity in $\dot{q}(q)$ shown in the bottom panel of Figure \ref{fig:qscatter}. 
 
\subsection{Eccentricity}\label{sec:eccentricity}
\begin{figure}
\includegraphics[width=\columnwidth]{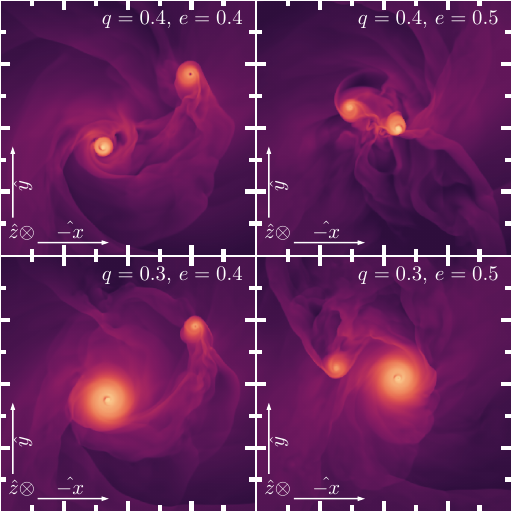}
\caption{Retrograde binaries with eccentricities of 0.4 and 0.5, and mass ratios of 0.3 and 0.4, centered on the center of mass of each binary. Although the minidisks of near-equal-mass retrograde binaries are largely stripped away, the primary in retrograde binaries with $q<0.4$ moves slowly enough to acquire sizable minidisks. At high enough eccentricities, the secondary can approach the primary closely enough to interact with its minidisk. When this occurs, binary eccentricities are damped, overcoming the eccentricity pumping from interactions with the circumbinary flow at apocenter. Major ticks are placed every $1.25\,a$ and minor ticks are placed every $0.675\,a$.}
\label{fig:slicePeri}
\end{figure}

We find that retrograde binaries that are both eccentric ($e\gtrsim0.4$) and low-mass-ratio ($q\lesssim0.4$) have their eccentricities damped. Retrograde binaries at higher mass ratios or lower eccentricities have their eccentricities increased over time, and prograde binaries of all eccentricities and mass ratios have their eccentricities damped, in accord with earlier three-dimensional studies of equal-mass binaries \citep{2024ApJ...964...61D,2024ApJ...970..107C}. The divergent behavior of low-mass-ratio retrograde binaries is due to the breakdown of the accretion morphology described earlier: at sufficiently low mass ratios ram pressure stripping becomes negligible within the Roche lobe of the primary, allowing substantial accretion disks to form. We illustrate the formation of a minidisk around the primary black hole in some eccentric retrograde binaries in Figure \ref{fig:slicePeri}. It is obvious that this change in behavior should occur as $q\rightarrow0$ as the orbital velocity of the primary about its center of mass tends towards zero and the difference between prograde and retrograde binaries becomes immaterial. At least for the accretion disk parameters in our current study the transition actually occurs at $q\lesssim0.4$, but the precise mass ratio likely depends on gas thermodynamics. 

It is useful to first consider the reason for eccentricity pumping in equal-mass retrograde binaries, as investigated thoroughly in \citet{2024ApJ...970..107C}. The dominant interaction between high-$q$ retrograde binaries and their environments is a drag force experienced as they plow headfirst through the medium surrounding them. At apocenter, the lever arm of a binary is quite long and the binary velocity is quite low, causing the drag force to sap away more angular momentum than energy from the binary, increasing its eccentricity. At higher mass ratios, neither object has much of a minidisk to speak of, making their mutual interaction near pericenter fairly negligible. At mass ratios low enough that the primary can sustain a minidisk, interactions near pericenter can damp eccentricities, as the dynamical conditions near pericenter reverse those near apocenter. However, for substantial eccentricity damping to occur near pericenter, the binary eccentricity must be large enough for the secondary to interact strongly with the accretion disk around the primary. Thus, eccentricity damping in retrograde binaries near pericenter is negligible at lower eccentricities ($e\lesssim0.3$) even when the primary sustains a minidisk, as in the $e=0.3, q=0.3$ case, for example. 

\begin{figure}
\includegraphics[width=\columnwidth]{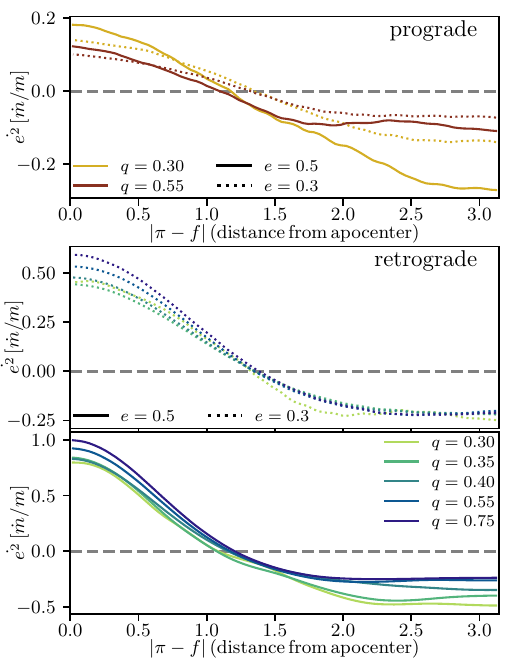}
\caption{ The average rate of change of $e^2$ as a function of orbital phase for a selection of retrograde and prograde binaries. In all cases eccentricities are damped near pericenter and pumped near apocenter, but the relative degree of pumping and damping depend on binary orientation, mass ratio, and eccentricity. Comparatively speaking, most retrograde binaries experience significantly enhanced eccentricity pumping near apocenter, as they experience a drag force. Higher-mass-ratio retrograde binaries experience relatively negligible eccentricity damping near pericenter, but those with low enough mass ratios to support circumprimary minidisks ($q=0.3$ and $q=0.35$ in the bottom panel) experience strong enough pericentric eccentricity damping to be driven towards attractors around $e\sim0.4$. Lower-eccentricity retrograde binaries do not approach closely enough at pericenter to experience additional drag, even at these lower mass ratios, as shown by comparison of the second and third panels near apocenter. }
\label{fig:efold}
\end{figure}

These trends in eccentricity damping and pumping as functions of orbital phase and mass ratio are illustrated more quantitatively in Figure \ref{fig:efold}, which plots the phase-averaged rate of eccentricity pumping folded about apocenter. In all prograde cases, eccentricity damping near pericenter overcomes the mild eccentricity pumping near apocenter. In most retrograde binaries, however, the relative strength of the apocentric contribution is greater than that of the pericentric contribution. Only retrograde binaries at low-$q$ and high-$e$ deviate from this trend, as the eccentricity damping at pericenter overwhelms the eccentricity pumping at apocenter. Generally, binary eccentricity has a much stronger effect on the magnitude of the phase-averaged rate of eccentricity pumping or damping than the binary mass ratio. Higher eccentricities make both apocentric and pericentric interactions between the binary and fluid more extreme, in terms of both lever arm extent and relative velocity. Although the lever arm of the interaction between the secondary and surrounding accretion flow is more extreme in lower mass-ratio binaries, this is counteracted by the lower mass of the secondary and thus the lower magnitude of its gravitational interactions with the surrounding medium.   

\section{Discussion}\label{sec:discussion}
Because most binaries formed within AGN disks should form eccentric and with $q<1$, the results presented above shed considerable light on the orbital evolution of these systems, which had previously been studied under the assumptions of either $e\sim0$ or $q\sim1$ \citep{2022MNRAS.517.1602L,2024ApJ...970..107C}, if not both \citep[e.g.,][]{2022ApJ...940..155D,2024ApJ...964...61D}. Thus, we can \textit{begin} to more realistically study how binary populations, once formed at $a\lesssim R_H \sim 10^6 R_g$, evolve towards merger or dissolution, which we pursue in Section \ref{sec:orbital}. However, our present study suffers from a number of limitations, which we discuss in Section \ref{sec:caveats}.

\subsection{Orbital Evolution}\label{sec:orbital}
\subsubsection{Spin and Mass-Ratio Evolution}
Population-level analyses of the GW events observed by LIGO and Virgo have shown modest but statistically significant signs of a negative correlation between binary mass ratios and the mass-weighed spin of each black hole projected onto the angular momentum of the binary orbit,$\chi_{\rm eff}$, as well as a paucity of binaries with $\chi_{\rm eff}<0$. \citep{2021ApJ...922L...5C,2023ApJ...958...13A}.
With the exception of low-mass-ratio retrograde binaries, we generally confirm for eccentric and unequal-mass binaries the result of \citet{2024ApJ...964...61D} that the minidisks formed around these embedded black holes typically form with the same rotational orientation as the orbit of the binary itself about its center of mass, driving such binaries towards spin-orbit alignment (and thus $\chi_{\rm eff}>0$). On the other hand, when the primaries of low-mass-ratio retrograde binaries are able to maintain accretion disks, the handedness of those accretion flows is the same as that of the larger-scale (prograde) accretion flow within the Hill sphere potentially driving those binaries towards spin-orbit misalignment. 

The degree of spin-orbit alignment to be expected from binaries in AGN disks is complicated by the change in accretion disk orientation at lower mass ratios. 
The spin-orbit evolution of higher-$q$ binaries may also be complicated as binaries are driven towards higher eccentricities, which will make them much easier to hydrodynamically realign with the AGN disk in a prograde sense, complicating their evolution in both spin and eccentricity. Although some of the trends in the spin and mass-ratio evolution show tantalizing resemblance to trends suggested by the population data, thorough population modeling will be necessary to make rigorous predictions for the spins and mass ratios of compact binary mergers in AGN disks.\footnote{A further complication is that accretion flows may severely limit black hole spins \citep[e.g.,][]{2023ApJ...954L..22R,2024ApJ...960...82L}.}

\subsubsection{The Possibility of Measurable Eccentricity in the LVK Band}
At mass ratios below $q\lesssim0.4$, the primary black hole in retrograde binaries can acquire a substantial minidisk, and at eccentricities $e\gtrsim0.4$ the secondary can approach closely enough at pericenter for the binary to experience eccentricity damping. Thus, while the eccentricities of prograde binaries always seem to be damped towards zero, retrograde binaries can reach a few different attractors in eccentricity depending on their mass ratio: lower-mass-ratio binaries are driven towards an equilibrium at about $e\sim0.3-0.4$, while higher-mass-ratio binaries are driven towards eccentricities of unity at which point GW emission can drive them to coalescence. Thus, near-equal-mass retrograde binaries can reach $e\sim1$ in less than a single mass $e$-folding time, whereas prograde and lower-mass-ratio retrograde binaries can take substantially longer to reach small semi-major axes.

\begin{figure}
\includegraphics[width=\columnwidth]{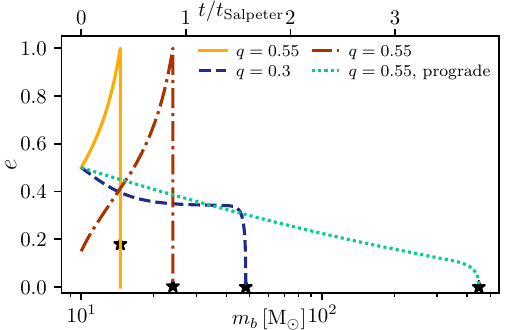}
\caption{Mock calculation of the evolution of a set of BHBs, initially $10\,M_\odot$ in mass, in eccentricity and mass, assuming accretion at the Eddington rate. Here we assume that each binary orbits at a distance of $10^4 \,{\rm AU}$ from a $10^{7}\,M_\odot$ SMBH with initial semi-major axes of $0.25 R_H$. We neglect changes in binary mass ratio over time, and evolve the binary according to a cubic spline interpolation of the results in Figure \ref{fig:scatter} in addition to the leading-order GW emission terms from \citet{1964PhRv..136.1224P}.}\label{fig:toy}
\end{figure}

In some cases, binaries having been driven to near-unity eccentricities may enter the LVK band with a detectable eccentricity; although the prospects for confident determination of binary eccentricities are limited by current detectors and waveform models \citep[e.g.,][]{2023MNRAS.519.5352R}, future detectors should more readily distinguish binary eccentricity \citep[e.g.,][]{2024MNRAS.528..833S,2025arXiv250312263A,2025arXiv250324084Z}. Furthermore, since binaries in AGN disks may accrete to fairly high masses, the lower frequencies to which future instruments aim to be sensitive should better probe populations of binaries within AGN disks.

To illustrate how binaries within AGN disks might evolve, Figure \ref{fig:toy} shows the evolution of a few binaries in mass and eccentricity subject to accretion at the Eddington rate (with a radiative efficiency of 10\%) and the time averaged rates of change of semi-major axis and eccentricity shown in Figure \ref{fig:scatter} and Table \ref{tab:orbit} along with the analogous orbit-averaged leading-order post-Newtonian energy and angular momentum loss terms from gravitational radiation (specifically, Equations 5.6 and 5.7 of \citealt{1964PhRv..136.1224P}). This toy model neglects change in binary mass ratio, which tends to change much more slowly than eccentricity, and the possibility of spin flips in retrograde binaries (see Section \ref{sec:caveats}), which might overestimate the maximum eccentricities reached by each binary. 

Nevertheless, the toy calculations of Figure \ref{fig:toy} illustrate a number of points that should qualitatively generalize: many prograde binaries, unless spurred on by other dynamical interactions, will accrete their way out of the LVK band or potentially outlive the AGN disk. Lower-mass-ratio retrograde binaries should also accrete substantially before merger, since they are unlikely to reach eccentricities capable of significantly accelerating their inspiral. Differences in the accretion rate onto the binary will primarily change the timescale required for each binary to reach the gravitational-wave dominated regime, shrinking or extending the time until merger; higher accretion rates have the potential to appreciably pump eccentricities until later in the inspiral, leading to slightly higher eccentricities near merger (and lower eccentricities at lower accretion rates). Both prograde binaries and low-mass-ratio retrograde binaries will likely enter the LVK band with $e\approx0$. Higher-mass-ratio binaries, on the other hand, may reach near-unity eccentricities without undergoing much accretion; they will therefore be more likely to reach the LVK band at lower masses and with potentially measurable eccentricity.

\subsection{Caveats}\label{sec:caveats}
The generality of our results is hindered by a few important assumptions made to facilitate our calculations: we have neglected the presence of magnetic fields and their influence on gas dynamics around the binary; we have assumed a globally isothermal equation of state; and we have ignored the effects of radiation. The lattermost assumption may be particularly damning, since embedded binaries accrete at nearly the Bondi rate, which is many orders of magnitude larger than the Eddington-limited rate, suggesting that outflows and other forms of feedback could significantly alter these accretion flows \citep[e.g.,][]{2014ApJ...796..106J,2023ApJ...946...26G}, although significant photon trapping and advection \citep[e.g.,][]{2012MNRAS.425.2892W,2022ApJ...934...58J} could leave the overall picture largely unchanged. Two-dimensional simulations suggest that non-isothermal thermodynamics can significantly affect the orbital evolution of embedded binaries, either by accelerating inspirals by orders of magnitude or be causing binaries to outspiral \citep{2023MNRAS.522.1881L}. It is yet unclear how non-isothermal gas dynamics would alter the results of three-dimensional calculations, but simulations of binary formation in AGN disks suggest that the nature of the large-scale accretion flow around the binary could be altered by outflows launched from the minidisks \citep{2025arXiv250214959W}. Magnetic fields might also modify the accretion flow by introducing large-scale turbulent inhomogeneities \citep[e.g.,][]{2024arXiv240905614M} or potentially by spurring jet formation. 

Unfortunately the expense of simulations employing more realistic physical approximations generally precludes them and thus our parameter survey must be interpreted with care. Most of the effects discussed above should lead to changes in the vertical structure of the accretion flow, likely leading to a mix of inflows and outflows without disrupting gas near the midplane as strongly. Presumably, the general trends with orbital orientation, mass ratio, and eccentricity should hold, but the numerical values measured from our simulations should be interpreted with these limitations in mind. 

Although the parameter survey of three-dimensional calculations presented herein is the most thorough yet conducted, it is still lacking in a few respects. For example, as retrograde binaries become more eccentric they should be more easily aligned with the disk in a prograde sense by out-of-plane torques \citep[e.g.,][]{2005MNRAS.363...49K,2014MNRAS.439.3476R}, and thus simulations of eccentric slightly-misaligned binaries would be of interest. Additionally, we found that retrograde binaries reach a fixed point in eccentricity when the secondary interacts with the minidisk of the primary (if existent). Since even the high-$q$ retrograde binaries maintain minidisks $\sim0.05a$ in size, this reasoning suggests that retrograde binary eccentricities might only be excited to values of $\sim0.95$. However, we cannot consider the size of the minidisks in our simulations to be physically realistic: the resolution tests in \citet{2024ApJ...964...61D} found that minidisk sizes decrease with smaller sink and gravitational softening radii, and two-dimensional simulations of retrograde binaries find no minidisks at all \citep{2022MNRAS.517.1602L}, suggesting that eccentricities might actually be pumped towards unity with sufficient resolution (and thus perhaps in reality). 

\section{Conclusions}\label{sec:conclusion}
Our present study has examined accretion onto eccentric and unequal-mass binaries embedded within AGN disks, and the orbital evolution thereof, by means of three-dimensional shearing-box hydrodynamical simulations. This survey has greatly extended the parameter space of binary orbital elements explored by high-resolution three-dimensional hydrodynamical simulations, and represents a crucial step towards moving from global predictions of binary formation rates to the expected merger rates and properties of GW sources. Additionally, we have identified --- at lower mass ratios in retrograde binaries --- radical departures in flow morphology and orbital evolution from the behavior observed at higher mass ratios. 

Prograde binaries tend to inspiral more slowly at lower mass ratios, with little dependence on eccentricity, and tend to be driven towards equal masses by accretion; prograde binary orbital evolution is fairly insensitive to mass ratio. High-mass-ratio retrograde binaries evolve similarly to equal-mass retrograde binaries \citep{2024ApJ...964...61D,2024ApJ...970..107C}, experiencing eccentricity excitation rather than damping and inspiralling due to hydrodynamical interactions more quickly than their prograde counterparts. However, at mass ratios $q\lesssim0.4$, substantial minidisks form around the primary in each binary, facilitating hydrodynamical interactions near pericenter that damp eccentricities, driving low-$q$ systems towards a moderate-$e$ attractor. Accretion also drives retrograde binaries towards more extreme mass ratios, albeit relatively slowly.

Our results suggest that prograde binaries and low-mass ratio binaries may accrete for multiple $e$-folding timescales at low eccentricities before reaching small enough separations for GWs to play a role in their dynamics. Higher-mass ratio retrograde binaries, however, might be fairly rapidly driven towards near-unity eccentricities, at which point GW emission near pericenter may drive their merger before much accretion can take place. Thus, under the dynamical influence of gas, binaries may reach the LVK band with relatively high masses, potentially within the upper mass gap, or lower masses and some residual eccentricity. 

\section*{Software}
\texttt{matplotlib} \citep{4160265}, \texttt{cmocean} \citep{cmocean}, \texttt{numpy} \citep{5725236}, \texttt{yt} \citep{2011ApJS..192....9T}, \texttt{Athena++} \citep{2020ApJS..249....4S}, \texttt{REBOUND} \citep{2012A&A...537A.128R,2015MNRAS.446.1424R}

\section*{Acknowledgments}
We gratefully acknowledge the support by LANL/LDRD under project number 20220087DR. This research used resources provided by the Los Alamos National Laboratory Institutional Computing Program, which is supported by the U.S. Department of Energy National Nuclear Security Administration under Contract No. 89233218CNA000001. The LA-UR number is LA-UR-25-24432.
Support for this work was provided by NASA through the NASA Hubble Fellowship grant \#HST-HF2-51553.001 awarded by the Space Telescope Science Institute, which is operated by the Association of Universities for Research in Astronomy, Inc., for NASA, under contract NAS5-26555.

\bibliographystyle{aasjournalnolink}
\bibliography{references}

\appendix
\section{Quantitative Orbital Evolution Results}\label{sec:appendix}
Table \ref{tab:orbit} lists the time-averaged orbital evolution results from our simulations. See Appendix A of \citet{2024ApJ...964...61D} for a description of how these quantities were calculated. Note that the orbital evolution contributions due to accretion also be calculated analytically if accretion disks form around each black hole prior to accretion: in this case, gas accretes on average with the same velocity as the object onto which it accretes, leaving the specific energy and specific angular momentum of the binary unchanged; the rate of change of the binary semi-major axis and eccentricity can thus be calculated analytically. In this limit, which is applicable to prograde binaries, accretion causes $\langle\dot{a}\rangle/a=-\dot{m}/m$ and $\dot{e}=0$. Thus, truncation errors in resolving accretion onto the black holes in our simulations result in orbital evolution inference errors on the order of a few per cent. Since retrograde binaries experience a strong headwind, the above reasoning does not generally apply, and deviations from this prediction instead quantify the importance of the headwind experience by each black hole.
\begin{deluxetable*}{cccccccccccccc}
\caption{Time-averaged values relevant to binary orbital evolution.}\label{tab:orbit}
\tablehead{
\colhead{$q$} & 
\colhead{$e$} & 
\colhead{$\langle\dot{m}\rangle/\dot{m}_B$} & 
\colhead{$\langle\dot{m}_2\rangle/\langle\dot{m}_1\rangle$} & 
\colhead{$\langle\dot{q}\rangle$} & 
\colhead{$\langle\dot{a}\rangle/\langle a \rangle$} & 
\colhead{$\langle\dot{a}_a\rangle/\langle a \rangle$} & 
\colhead{$\langle\dot{a}_g\rangle/\langle a \rangle$} & 
\colhead{$\langle\dot{e^2}\rangle$} & 
\colhead{$\langle\dot{e^2}_a\rangle$} & 
\colhead{$\langle\dot{e^2}_g\rangle$} & 
\colhead{$\langle\dot{\varpi}\rangle$} & 
\colhead{$\langle\dot{\varpi}_a\rangle$} & 
\colhead{$\langle\dot{\varpi}\rangle$} 
}
\startdata
Prograde & & & & & & & & & & & & & \\ \hline
0.3 & 0.0 & 0.229 & 0.327 & 0.029 & -3.236 & -1.048 & -2.188 & 0.027 & 0.019 & 0.008 & N/A & N/A & N/A \\
0.3 & 0.3 & 0.228 & 0.357 & 0.060 & -3.664 & -1.033 & -2.631 & -0.429 & 0.036 & -0.465 & 0.723 & -0.160 & 0.883 \\
0.3 & 0.5 & 0.240 & 0.455 & 0.160 & -3.589 & -0.999 & -2.590 & -0.638 & 0.066 & -0.704 & 1.639 & -0.105 & 1.744 \\
0.55 & 0.0 & 0.255 & 0.691 & 0.159 & -2.179 & -1.101 & -1.078 & 0.010 & 0.020 & -0.010 & N/A & N/A & N/A \\
0.55 & 0.3 & 0.261 & 0.768 & 0.241 & -2.287 & -1.082 & -1.204 & -0.227 & 0.049 & -0.276 & 0.464 & -0.283 & 0.747 \\
0.55 & 0.5 & 0.277 & 0.892 & 0.375 & -1.902 & -0.983 & -0.919 & -0.256 & 0.083 & -0.339 & 1.237 & -0.183 & 1.420 \\
\hline
Retrograde  & & & & & & & & & & & & & \\ \hline
0.3 & 0.0 & 0.333 & 0.101 & -0.379 & -5.058 & -0.851 & -4.206 & 0.107 & 0.017 & 0.090 & N/A & N/A & N/A \\
0.3 & 0.3 & 0.328 & 0.132 & -0.305 & -5.854 & -0.903 & -4.951 & 0.199 & 0.031 & 0.168 & 1.796 & 0.067 & 1.729 \\
0.3 & 0.4 & 0.330 & 0.134 & -0.302 & -6.589 & -0.876 & -5.713 & -0.204 & 0.034 & -0.238 & 1.038 & 0.094 & 0.945 \\
0.3 & 0.5 & 0.314 & 0.130 & -0.295 & -7.928 & -0.940 & -6.988 & -0.751 & -0.010 & -0.741 & 0.226 & 0.079 & 0.148 \\
0.35 & 0.0 & 0.337 & 0.134 & -0.419 & -5.458 & -0.866 & -4.591 & 0.115 & 0.017 & 0.098 & N/A & N/A & N/A \\
0.35 & 0.3 & 0.335 & 0.138 & -0.407 & -5.457 & -0.879 & -4.579 & 0.261 & 0.005 & 0.256 & 0.866 & 0.042 & 0.823 \\
0.35 & 0.4 & 0.335 & 0.144 & -0.392 & -5.643 & -0.876 & -4.767 & 0.223 & -0.013 & 0.236 & 0.807 & 0.087 & 0.720 \\
0.35 & 0.5 & 0.327 & 0.161 & -0.346 & -7.771 & -0.946 & -6.824 & -0.515 & -0.006 & -0.509 & 0.331 & 0.084 & 0.246 \\
0.4 & 0.0 & 0.339 & 0.176 & -0.437 & -5.742 & -0.869 & -4.873 & 0.123 & 0.018 & 0.105 & N/A & N/A & N/A \\
0.4 & 0.3 & 0.338 & 0.178 & -0.432 & -5.708 & -0.870 & -4.838 & 0.309 & 0.006 & 0.302 & 0.723 & -0.070 & 0.792 \\
0.4 & 0.4 & 0.337 & 0.183 & -0.416 & -5.681 & -0.865 & -4.816 & 0.369 & -0.012 & 0.381 & 0.682 & -0.014 & 0.696 \\
0.4 & 0.5 & 0.343 & 0.191 & -0.408 & -6.313 & -0.962 & -5.351 & 0.095 & -0.073 & 0.168 & 0.488 & -0.006 & 0.495 \\
0.55 & 0.0 & 0.341 & 0.368 & -0.339 & -6.484 & -0.848 & -5.636 & 0.140 & 0.019 & 0.121 & N/A & N/A & N/A \\
0.55 & 0.3 & 0.341 & 0.375 & -0.325 & -6.197 & -0.833 & -5.364 & 0.457 & 0.017 & 0.441 & 1.099 & -0.067 & 1.166 \\
0.55 & 0.4 & 0.340 & 0.374 & -0.325 & -6.160 & -0.828 & -5.332 & 0.548 & 0.002 & 0.546 & 0.869 & -0.041 & 0.910 \\
0.55 & 0.5 & 0.333 & 0.350 & -0.369 & -6.393 & -0.943 & -5.450 & 0.482 & -0.067 & 0.549 & 0.635 & -0.043 & 0.678 \\
0.75 & 0.0 & 0.340 & 0.722 & -0.046 & -6.915 & -0.874 & -6.041 & 0.154 & 0.019 & 0.135 & N/A & N/A & N/A \\
0.75 & 0.3 & 0.340 & 0.703 & -0.079 & -6.493 & -0.844 & -5.650 & 0.605 & 0.017 & 0.589 & 1.050 & -0.027 & 1.077 \\
0.75 & 0.4 & 0.341 & 0.696 & -0.091 & -6.286 & -0.815 & -5.472 & 0.777 & 0.009 & 0.768 & 0.967 & 0.007 & 0.959 \\
0.75 & 0.5 & 0.347 & 0.644 & -0.190 & -6.252 & -0.881 & -5.370 & 0.799 & -0.051 & 0.850 & 0.709 & 0.002 & 0.707 \\
\enddata
\tablecomments{Note that with the exception of the accretion rates themselves, each reported time derivative as been normalized by $\langle\dot{m}\rangle/m$. Also note that the accretion rates onto the binary are very close to that predicted by minor tidal truncation of the accretion flow, which predicts $\dot{m}/\dot{m}_B\sim(R_H/R_B)^2=0.2713$. }
\end{deluxetable*}

\end{document}